\documentclass{article}

\usepackage{arxiv}

\usepackage[utf8]{inputenc} 
\usepackage[T1]{fontenc}    
\usepackage{hyperref}       
\usepackage{url}            
\usepackage{booktabs}       
\usepackage{amsfonts}       
\usepackage{amssymb}
\usepackage{amsmath}
\usepackage{microtype}      
\usepackage{graphicx}
\usepackage{authblk}
\title{The Elastic Properties and Lattice Dynamics for Selected 211 MAX Phases: A DFT Study}

\author[1]{G. K. Arusei$^{*}$}
\author[2]{M. Chepkoech}
\author[3]{G. O. Amolo}
\author[4]{N. Makau}
\affil[1,4]{Department of Physics, University of Eldoret, P.O Box 1125, Eldoret (Kenya)}
\affil[2]{University of Kabianga (UoK), P.O.Box 2030, 20200 Kericho, Kenya }
\affil[3]{School of Physics and Earth Sciences, The Technical University of Kenya, P.O. Box 52428-00200, Nairobi, Kenya}
\affil[1]{Email id: aruseig@yahoo.com}
\date{}
\renewcommand{\shorttitle}{\textit{arXiv} Template}

\hypersetup{
pdftitle={A template for the arxiv style},
pdfauthor={G. K. Arusei et al.},
}

\begin{document}
\maketitle

\begin{abstract}
{The elastic properties and lattice dynamics of Ti$_2$AlC, Ti$_2$AlN, Ti$_2$GaC, Ti$_2$GaN, Ti$_2$PbC, Ti$_2$CdC and Ti$_2$SnC have been investigated using the density functional theory within the generalized gradient approximations as expressed in Quantum Espresso and VASP codes. The obtained lattice parameters are in agreement with the previous theoretical research and available experimental data. The elastic properties of the MAX phases under study have been calculated. The values of elastic anisotropy, Young's modulus, Poisson ratio and shear modulus reveal that the compounds are stable and ductile and that Ti$_2$PbC and Ti$_2$CdC are more stable than the other considered compounds. Thus, the seven compounds may be useful for industrial applications. The calculated phonon spectra confirm that the studied MAX phases are dynamically stable because of the absence of imaginary phonon modes. The temperature dependent lattice thermal conductivity of MAX phases have been determined using the Debye theory as outlined by Slack. The obtained $\kappa_{ph}$ for Ti$_2$AlC at 1300 K agree with experimental findings within 9$\%$. The estimated minimum thermal conductivities ($\kappa_{min}$) of the MAX phases obtained by using empirical formula suggested by Clarke show that Ti$_2$PbC possesses the lowest minimum thermal conductivity. 
}
\end{abstract}

\section{Introduction}
MAX phases materials presented in this paper (Ti$_2$AlC, Ti$_2$AlN, Ti$_2$GaC, Ti$_2$GaN, Ti$_2$PbC, Ti$_2$CdC and Ti$_2$SnC) are ternary carbides and nitrides materials with the general formula M$_{n+1}$
X$_n$(MAX) where $n$ = 1, M is an early transition metal, A is an A-group element (mostly IIIA and IVA) and X is either C or N-represent a new class of thermodynamically stable nano-laminated solids \cite{Bor} with unique combination of ceramic and metallic properties. The M$_{n+1}$ AX$_n$ phases crystallize in the hexagonal structure belonging to the space group of P63/mmc. MAX phases mimic ceramics in that they are stiff, resistant to oxidation, and remain strong at temperatures beyond 1400C. The metal-like properties of MAX phases manifest themselves in their machinability, resistance to thermal shock, high damage tolerance, and electrical and thermal conductivity \cite{Bor,Bor1}. These unique combination of properties suits as structural materials for demanding environmental operations like increasing the efficiency of engines requires operation at much higher temperatures than what is allowed by today's materials.
\par Up to now, lack of large MAX single crystals has made it very hard to determine experimentally the elastic constants of these compounds. However, the ab initio results are available. Research efforts have been made to study the physical properties of M$_2$AX phases both theoretically and experimentally \cite{bar,Fin,Bars,Erk,Nag,Hor,Yoo,Al,Barso,Ali} and a few of these materials have been fully investigated. Among the compounds, there are few studies on technological importance on M$_2$InC (M = Zr, Hf and Ta) MAX phases \cite{Barso,gupta,mano,barsoum,med,he}. \par A lot of work has been carried out with Al as an A element and C as X element in MAX-series \cite{e11,chandra,e1}. Moreover, the structural, electronic and elastic properties of M$_2$GaN compounds have been done by Bouhemadou \cite{bouh}. Dhakal et al. \cite{chandra} have also calculated the lattice thermal conductivity of 551 MAX phase compounds using the Debye theory as outlined by Slack model \cite{slack}.
\par Theoretical calculations on the mechanical properties, lattice dynamics and thermal conductivity of the 211 MAX phases under study are limited and thus in this work, first principle density functional theory calculations will be done to examine the elastic properties of the selected materials with the aim of identifying optimum compositions that combine desirable machinability with high stiffness. In addition, the phonons and approximated thermal conductivity studies will be performed on these compounds and compared with the available theoretical and experimental data.
\section{Computational details}
The structural and mechanical properties of the MAX phases under study were determined by using density functional theory \cite{Kohn,dft1} as implemented in the Quantum Espresso (QE) \cite{qe,qe1,qe2} and VASP \cite{viena} codes. The exchange-correlation functional is approximated using the Generalized Gradient Approximation (GGA) as proposed by Perdew, Burke and Ernzerhof \cite{Perdew}. The core electrons were described by the projector augmented wave method (PAW) \cite{paw,paw1} (in VASP) and ultrasoft pseudopotentials \cite{u} (in QE).  A plane-wave kinetic cut-off energies of 60 Ry (for Ti$_2$SnC), 50 Ry (for Ti$_2A$lC, Ti$_2$AlN, Ti$_2$GaC and Ti$_2$GaN), 40 Ry (for Ti$_2$PbC) and 30 Ry (for Ti$_2$CdC) were determined through a careful convergence scheme. In VASP code, a kinetic cut-off of 520 eV was used to expand the electronic wave functions. The first Brillouin zone was sampled using a 12$\times$12$\times$10 Monkhorst-Pack k-point grid \cite{monk}. \par In order to obtain reliable results, a highly accurate computational method, THERMO-PW method \cite{Sund} was employed in this study to calculate the elastic constants as implemented in the QE code using the three accepted elastic stability for selected MAX Phases in this study \cite{maxx,Sund}. This method has proven successful in theoretical studies of elastic properties on various materials \cite{maxx,Sund}. In the case of VASP code, the elastic constants were calculated by the efficient stress-strain method \cite{elast}. The Phonopy code \cite{phonopy} was used to calculate the lattice dynamics of the MAX phases at the level of Harmonic approximations by using the finite displacement method. The second-order harmonic force constant were computed using a 2$\times$2$\times$2 supercells consisting of 64 atoms. The temperature dependent lattice phonon thermal conductivity of MAX phases are determined using Slack’s equation \cite{slack}.
\section{Results and discussions}
\subsection{Structural properties}
\begin{figure}[!ht]
\centering
\includegraphics[width=10.25pc]{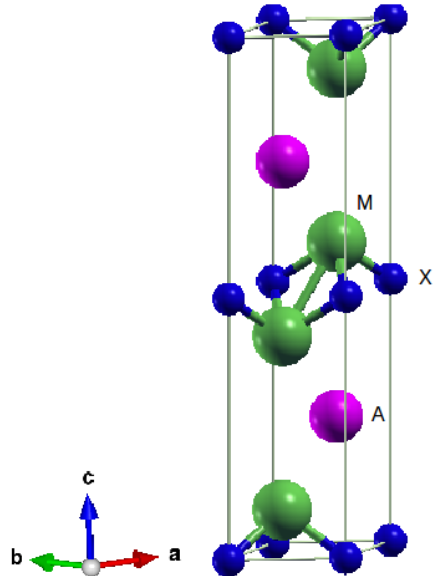}
\caption{\label{Fig:m}Crystal structure of the M$_2$AX phase.}
\end{figure}

M$_{2}$AX crystallizes in hexagonal structure with the space group $\textit{P63mmc}$ no. 194. The unit cell structure of M$_2$AX compounds is depicted in Fig. \ref{Fig:m}. The unit cell composes of 2 formula units with 8 atoms.
The equilibrium lattice parameters, such as the lattice constants and equilibrium cell volume of the MAX phases are calculated and are tabulated in Table \ref{tab:11}. Our calculated results are compared with other reported results and available experimental data as listed in Table \ref{tab:11}. The obtained lattice parameters are in good agreement with other theoretical data  as listed in Table \ref{tab:11}. The calculated lattice constants deviate from the experimental data by less than 2$\%$. 

\begin{table*}[!h]
\caption{\label{tab:11}The calculated equilibrium lattice parameters, $a$ and $c$, (in \AA), equilibrium cell volume, $V_0$, (in \AA$^{3}$) of considered MAX phases obtained using VASP and QE codes compared with available theoretical and experimental data.}
\centering
 \begin{tabular}{c c c c c} 
 \hline
&$a$& $b$&$V_0$&Reference\\ 
 \hline
 Ti$_2$AlN&2.99 - 3.07&13.51 - 13.65&106 - 110.16&This Calc.\\ 
          &2.961&13.652&106&\cite{bouh}\\
           &2.989&13.614&105.33& \cite{Md}\\
           &2.98&13.68&105.21& Expt. \cite{calc}\\
            &3.01&13.70&107.49& \cite{calc}\\ 
          \hline
 Ti$_2$AlC&3.07 - 3.10&13.74 - 13.82&115.57 - 112.08&This Calc.\\
 \hline
  &3.051&13.637&109.93& \cite{Md}\\
  &3.04&13.59&108.77& Expt. \cite{calc}\\
   &3.08&13.77&113.13& \cite{calc}\\ 
          \hline
 Ti$_2$GaC&3.07 - 3.08&13.46 - 13.51&110.16 - 110.56&This Calc.\\ 
  &3.07&13.52&110.35& \cite{Md}\\
          \hline
 Ti$_2$GaN&3.02 - 3.07&3.352 - 13.51&105.12 - 110.16&This Calc.\\ 
          &2.961&13.021&98.87&LDA \cite{bouh}\\
          &3.018&13.318&105.07&GGA \cite{bouh}\\
                     &3.00&13.3&103.7& Expt. \cite{barsoum}\\
          \hline
  Ti$_2$SnC&3.18&13.80 - 13.97&120.48 - 121.96&This Calc.\\ 
   &3.163&13.679&118.52& \cite{Md}\\
   &3.1626&13.679&-& Expt. \cite{exp}\\
   &3.164&13.675&-& Expt. \cite{exp1}\\
   &3.186&13.63&-& Expt. \cite{exp2}\\
          \hline
 Ti$_2$CdC&3.07 - 3.11&14.118 - 14.53&115.16 - 121.36&This Calc.\\
  &3.1&14.41&10119.92& \cite{Md}\\ 
   &3.099&14.41&119.85& Expt. \cite{exp3}\\
          \hline
 Ti$_2$PbC&3.23 - 3.28&13.78 - 14.01&126.48 - 128.47&This Calc.\\
  &3.20&13.81&122.46&\cite{Md}\\
  &3.222&13.99&125.78&Expt. \cite{exp3}\\
  &3.201&13.78&122.28&Expt. \cite{exp4}\\
   \hline\hline
 \end{tabular}
\end{table*}
\subsection{Mechanical properties}
\par The mechanical behavior of solids depends upon their elastic constants. The elastic constants provide information on the stability, stiffness, brittleness, ductility, and elastic anisotropy of a given material. The MAX phases possess hexagonal crystal structures \cite{Md,Ani} and thus have six elastic constants: $C_{11}$, $C_{12}$, $C_{13}$, $C_{33}$, $C_{44}$ and $C_{66}$. The first five are independent elastic constants whereas the last one takes into account: $C_{66}$ = ($C_{11}$-$C_{12}$)/2. \par A mechanically stable MAX compound should obey the following stability conditions \cite{stability1,stability2,stability3}: 

$$C_{11}>0, C_{33}>0, C_{44}>0, (C_{12}-C_{12})>0, \text{   and  }$$ 
$$(C_{11}+C_{12})C_{33}>2(C_{12})^2.$$
\par The predicted elastic properties for considered materials MAX phases are tabulated in Table \ref{tab:elas} together with the available theoretical data \cite{e11,e1,e2,e3,e4,e5}. From Table \ref{tab:elas1}, it can be seen that our estimated elastic constants obey the above mentioned Born stability conditions \cite{stability1,stability2,stability3} and thus confirms the studied MAX compounds are mechanically stable. Our calculated elastic constants for studied MAX phases are consistent with reported theoretical data \cite{e11,e1,e2,e3,e4,e5}.
\par For all the materials $C_{33}> C_{11}$(see Table \ref{tab:elas1}), implying that the compounds are more incompressible along the $c$-direction than along the $a$-direction. It also indicates that the compounds are elastically anisotropic. 
 \par The elastic modulus ($B, G, E$, and $\nu$) have been calculated from the elastic constants, $C_{ij}$ using the Voigt-Reuss-Hill formula \cite{VRH} and listed in Table \ref{tab:elas}. The bulk modulus provides a measure of resistance to volume change and the average bond strength in a material. The higher the bulk modulus the greater the bond strength. It is evident that Ti$_2$CdC possesses a smaller value of bulk modulus than those of other considered materials, and thus it possesses the lowest bond strength. Shear modulus plays a dominant role in predicting the hardness rather than bulk modulus. From Table \ref{tab:elas} the shear modulus of Ti$_2$AlN is found to be  higher than those of other materials under study, and as result, it is expected that Ti$_2$AlN should be harder compared to the other compounds.
\begin{table*}[!h]
\caption{\label{tab:elas1}Calculated elastic constants $C_{ij}$ (Gpa) for Ti$_2$AlC, Ti$_2$AlN, Ti$_2$GaC, Ti$_2$GaN, Ti$_2$PbC, Ti$_2$CdC and Ti$_2$SnC MAX phases obtained using VASP and QE codes compared with available theoretical and experimental data.}
\centering
 \begin{tabular}{c c c c c c c c} 
 \hline
&$C_{11}$&$C_{12}$& $C_{13}$&$C_{33}$&$C_{44}$&$A$&Reference\\ 
 \hline
 Ti$_2$AlN&272.1 - 308.4&56.8 - 75.3&86.9 - 89.5&271.3 - 287.2&121.1 - 118.6&1.14 - 1.26&This Calc\\         
    &342&56&96&283&123&1.14&\cite{e1}\\      
     &309&66&91&280&125&1.23&\cite{e11}\\       
\hline          
 Ti$_2$AlC&274.7 - 297.3&50.4 - 63.6&49.3 - 59&246.8 - 262&101.7 - 110&0.96 - 0.98&This Calc.\\ 
      &301&59&55&278&113&0.96&\cite{e2}\\      
      &302&62&61&269&109&0.97&\cite{e11}\\     
\hline  
Ti$_2$GaC&275.9 - 306.4&59.5 - 65.6&46.5 - 60.6&225.5 - 265.9&89.8 - 114&0.88 - 1.01&This Calc.\\ 
                   &314&66&59&272&122&1.04&\cite{e4}\\ 
                   \hline
 Ti$_2$GaN&267.9 - 293&70.3 - 89.9&85.4 - 91.6&249.5 - 272.9&107.5 - 113&1.18 - 1.24&This Calc.\\
  &338$\pm$2&97$\pm$2&111$\pm$0.8&312$\pm$3&136$\pm$1&1.27&LDA\cite{bouh}\\
 &296$\pm$1&84$\pm$0.7&92$\pm$0.4&275$\pm$3&119$\pm$0.5&1.23&GGA\cite{bouh}\\
         \hline  
 Ti$_2$SnC&228.0 - 265.5&74.2 -  78.2&47.7 - 71.2&202.7 - 263&74.0 - 91&0.88 - 0.94&This Calc.\\
                    &260&78&70&254&93&0.99&\cite{e3}\\      
          
\hline  
 Ti$_2$CdC&226.5 - 251.1&59.2 - 73.2&43.8 - 44.9&183.6 - 203.6&29.8 - 79.4&0.33 - 0.98&This Calc.\\ 
                    &258&68&46&205&33&0.36&\cite{e5}\\      
\hline
 Ti$_2$PbC&202.5 - 239.1&81.9 - 90.6&49.7 - 51.4&195.7 - 213.2&60.8 - 65.1&0.75 - 0.81&This Calc.\\
                   &235&90&53&211&66&0.78&\cite{e3}\\  
  \hline\hline
 \end{tabular}
\end{table*}
\begin{table*}[!h]
\caption{\label{tab:elas}Calculated elastic moduli $B$, $G$, $E$ (Gpa) and $\nu$ for Ti$_2$AlC, Ti$_2$AlN, Ti$_2$GaC, Ti$_2$GaN, Ti$_2$PbC, Ti$_2$CdC and Ti$_2$SnC MAX phases obtained using VASP and QE codes compared with available theoretical and experimental data.}
\centering
 \begin{tabular}{c c c c c c c c} 
 \hline
&$B$&$G$&$E$&$B/C_{44}$&$B/G$&$\nu$&Reference\\ 
 \hline
 Ti$_2$AlN&141.7 - 155.1&108.2 - 126&258.6 - 277.3 &1.17 - 1.31&1.23 - 1.31&0.19 - 0.2&This Calc.\\
             &162.55&126&300&1.32&1.29&0.192&\cite{e1}\\      
     &177&127&307&1.42&1.39&-&\cite{e11}\\       
\hline          
 Ti$_2$AlC &121.4 - 136.1&106.1 - 115&246.5 - 267&1.19 - 1.24&1.14 - 1.18&0.16 - 0.17&This Calc.\\
           &135.33&118&277&1.20&1.15&0.164&\cite{e2}\\      
      &138&113&267&1.27&1.22&-&\cite{e11}\\     
\hline  
 Ti$_2$GaC&140.8 - 153.6&98.6 - 108.2&239.8 - 263.5&1.31 - 1.35&1.42- 1.43&0.17 - 0.22&This Calc.\\

           &140.88&121&283&1.20&1.16&0.166&\cite{e4}\\  
           \hline  
 Ti$_2$GaN&119.5 - 138.8&98.8 - 110.5&232.3 - 262&1.23 - 1.33&1.21 - 1.26&0.18 - 0.19&This Calc.\\
           & 181&122&300&1.33&1.48&0.22& LDA \cite{bouh}\\   
            & 156&108&264&1.44&1.32&0.218& GGA \cite{bouh}  \\
\hline  
 Ti$_2$SnC&110.3 - 137.2&77.4 - 93&188.2 - 228&1.49 - 1.51&1.42 - 1.48&0.22&This Calc.\\
          &134.44&93&226&1.45&1.45&0.218&\cite{e3}\\      
         \hline  
 Ti$_2$CdC&102.6 - 113.5&61.5 - 65&148.2 - 153.6&2.57 - 3.81&1.67 - 1.75&0.24 - 0.25&This Calc.\\
          &115.66&69.6&174&3.51&1.66&0.249&\cite{e5}\\      
\hline
 Ti$_2$PbC&106.7 - 118.9&63.9 - 74&159.8 - 183.9 &1.75 - 1.83&1.61 - 1.67&0.24 - 0.25&This Calc.\\
          &119.22&73.2&182&1.81&1.63&0.245&\cite{e3}\\  
  \hline\hline
  \end{tabular}
     \end{table*}
 \par Young's modulus ($E$) is used to measure of stiffness of a material, i.e., the larger the value of $E$, the stiffer the material and vice versa. The $E$ values of the materials under study decrease in the order: Ti$_2$AlN $>$ Ti$_2$AlC $>$ Ti$_2$GaN $>$ Ti$_2$GaC $>$ Ti$_2$SnC $>$ Ti$_2$PbC $>$ Ti$_2$CdC. Thus, Ti$_2$AlN is stiffest (see Table \ref{tab:elas}).
\par Poisson's ratio plays another important role in assessing the nature of chemical bonding in solid materials \cite{poison}. The values of Poisson's ratio for pure covalent and ionic crystal are respectively, 0.1 and 0.25.  The Poisson's ratio for the studied materials lies between these two characteristic values, indicating that the compounds possess mixture of covalent and ionic bonding.
\par The elastic anisotropy factor $A$ determines how the elastic properties of a solid are dependent on the direction of the stress. Moreover, the elastic anisotropy is connected with the thermal expansion and the crystal micro-cracks \cite{anisotropy}. For the MAX phase systems that are hexagonal, the elastic anisotropy factor is calculated from the equation $A = 4C_{44}/(C_{11} + C_{33} - 2C_{13})$, and if $A$ = 1, the crystal is isotropic while any value smaller or lager than 1 indicates anisotropy. The calculated shear anisotropic factors for the MAX phase compounds are listed in Table \ref{tab:elas1}. The results from Table \ref{tab:elas1} characterize all the studied MAX phases as being elastically anisotropic. Moreover, Ti$_2$CdC is considered more anisotropic compared to the other studied compounds. This anisotropy originates from the crystal structure of the studied compounds.
 \par Other important macroscopic properties that depend on the elastic constants are the machinability and ductility \cite{mach,mach1}. Machinability is defined as the ratio of bulk modulus to $C_{44}$ ($B$/$C_{44}$), while ductility is defined as the ratio between bulk modulus and shear modulus, $B$/$G$. From Table \ref{tab:elas}, it is clear that Ti$_2$CdC possesses the largest value of $B$/$C_{44}$ and hence more machinable than the rest of the studied MAX phase compounds. The ratio of the bulk modulus ($B$) to $C_{44}$ may be interpreted as a measure of plasticity \cite{plas}. If the values of $B/C_{44}$ are large, it indicate that the material possesses excellent lubricating properties. In terms of the ductility ($B/G$)  \cite{pugh}, it is found that Ti$_2$PbC and Ti$_2$CdC are the most ductile material and Ti$_2$AlC is the least ductile (see Table \ref{tab:elas}).
\subsection{Debye Temperature and Thermal conductivity}
 \par The Debye temperature is an important fundamental parameter which is closely related  to many physical properties such as specific heat, melting temperatures, thermal conductivity, thermal expansion and lattice vibration. The Debye temperature, $\theta_D$, can be extracted from elastic constants data using by the formula \cite{debye},
$$\theta_D = \frac{h}{k_B}\bigg(\frac{3nN_A\rho}{4{\pi}M}\bigg)^{\frac{1}{3}}v_m.$$
Here $h$ is Planck's constant; $k_B$, Boltzmann's constant; $N_A$, Avogadro's number; $\rho$, the density of a material; $M$, molecular weight; $q$, the number of atoms in a molecule and $v_m$, is the averaged sound velocity of the system and is obtained by $$v_m = \bigg[{\bigg(\frac{2}{v^{3}_t} + \frac{2}{v^{3}_l}\bigg)}/3\bigg]^{-\frac{1}{3}},$$
where $v_t$ and $v_l$ are the longitudinal and transverse sound velocities. $v_t$ and $v_l$ can be determined using the bulk and shear moduli of the materials from the Navier's equations:
 $$v_l = \sqrt{\frac{3B+4G}{3\rho}} \text{  and } v_t = \sqrt{\frac{G}{\rho}},$$
where $B$, $G$ and $\rho$ are respectively the bulk modulus, shear modulus and the density of the polycrystalline solid. The results for sound velocities, densities and $\theta_D$ are summarized in Table \ref{tab:elas11}. A careful look at Table \ref{tab:elas11}, shows that the Ti$_2$AlN has the highest Debye temperature while Ti$_2$PbC has the smallest. Higher Debye temperature corresponds to the better thermal conductivity of the compounds. Thus, Ti$_2$AlN is thermally more conductive in the studied 211 MAX phases. The sound velocities of the considered MAX phases follow the trend: Ti$_2$AlN $<$ Ti$_2$AlC $<$ Ti$_2$GaC $<$ Ti$_2$GaN Ti$_2$SnC $<$ Ti$_2$CdC $<$ Ti$_2$PbC. This is expected as the bulk and shear modulus of these materials decrease in the same sequence.

\begin{table*}[!h]
\caption{\label{tab:elas11}The estimated volumic density, $\rho$, (in g/m$^3$), sound velocities ($v_t$, $v_l$ and $v_m$ in km/s) and Debye temperature ($\theta_D$, in K) for Ti$_2$AlC, Ti$_2$AlN, Ti$_2$GaC, Ti$_2$GaN, Ti$_2$PbC, Ti$_2$CdC and Ti$_2$SnC MAX phases obtained using VASP and QE codes compared with available theoretical and experimental data.}
\centering
 \begin{tabular}{c c c c c c c c c c c c c} 
 \hline
&$\rho$&$v_t$& $v_l$&$v_m$&$\theta_D$&$T_m$&Reference\\ 
 \hline
 Ti$_2$AlN&4.12 - 4.28 &5229 - 5425&8231 - 8312&5564 - 5747&703 - 747&1574 - 1710&This Calc.\\ 
          \hline
 Ti$_2$AlC&3.87 - 3.99&5096 - 5368&8226 - 8285&5544 - 5616&700 - 727&1548 - 1639&This Calc.\\ 
          & -&-&-&-&716&-&\cite{chandra}  \\
          \hline
 Ti$_2$GaC&5.35 - 5.33&4382 - 4553 &6987 - 7327&4817 - 5019&590 - 623&1520 - 1672&This Calc\\ 
          \hline
 Ti$_2$GaN&5.41 - 5.67&4239 - 4364&7045 - 7272&4683 - 4828&585 - 610&1532 - 1642&This Calc.\\ 
           & 6.03&4506&7552&4988&642&-&LDA \cite{bouh}  \\
            & 5.67&4369&7275&4869&610&-&GGA \cite{bouh}  \\
          \hline
  Ti$_2$SnC&6.16 - 6.24&3601 - 3861&5980 - 6470&3861 - 3980&473 - 515&1342 - 1545&This Calc.\\
          \hline
 Ti$_2$CdC&6.35 - 6.02&3241 - 3286&5615 - 5766&3551 - 3652&422 - 439&1309 - 1413&This Calc.\\ 
          \hline
 Ti$_2$PbC&8.14 - 8.27&2841 - 2991&4922 - 5129&3151 - 3318&369 - 393&1255 - 1391&This Calc.\\
   \hline\hline
 \end{tabular}
\end{table*}
\par The melting temperatures ($T_m$) of hexagonal crystal can be calculated from the elastic constants data by using the empirical formula developed by Fine et al. \cite{fine}:
\begin{equation}
T_m = 354 + 1.5(2C_{11} + C_{33}).
\end{equation}
The calculated melting temperatures of MAX phases is tabulated in Table \ref{tab:elas11}. The high melting temperatures displayed by these MAX phase compounds are encouraging for their application as high temperature
structural materials.
\par Clarke \cite{clarke} proposed that a material at high temperature possesses a minimum thermal conductivity. The minimum thermal conductivity signifies the theoretical lower limit of intrinsic thermal conductivity of a crystal. It can be calculated from the average sound velocity using \cite{clarke}:
\begin{equation}
\kappa_{min}=k_Bv_m\big(\frac{nN_A\rho}{M}\big),
\end{equation}
where $k_B$, $v_m$, $N_A$, $\rho$ are respectively, Boltzmann's constant, average sound velocity, Avogadro's constant and density of a crystal. The Clarke's formula \cite{clarke} is useful in determining the temperature independent minimum thermal conductivity ($\kappa_{min}$).
 The calculated minimum thermal conductivity of 211 MAX phases are tabulated in Table \ref{tab:kmin}. It is clear from Table \ref{tab:kmin} that the minimum thermal conductivity is directly proportional to the average sound velocity in crystalline solids. As the average sound velocity of Ti$_2$PbC is lowest, the minimum thermal conductivity of this material is also lowest. 
\par In this study, the lattice thermal conductivities of the MAX phases were estimated using the Slack's model \cite{slack} and compared with the available experimental data. Slack's model is most efficient method for determining the lattice thermal conductivity for ceramics like MAX phases. According to Slack's model \cite{slack} , the temperature dependent lattice thermal conductivity, $\kappa_{ph}$ of 321 MAX phases can be evaluated using the empirical formula \cite{slack}
\begin{equation}
\kappa_{ph}=R\frac{M_{av}\theta^3_D\delta}{\gamma^2n^{2/3}T}
\end{equation}
Here $M_{av}$ is the average atomic weight (in kg/mol) in a molecule, $\theta_D$ is the Debye temperature (in K), $\delta$ is the cubic root of average atomic volume, $n$ is number of atoms per unit cell, $T$ is the absolute
temperature, $\gamma$ is the Gr\"uneisen parameter derived from Poison's ratio ($\nu$) and $R$ is a coefficient (in W-mol/kg/m$^2$/K$^3 $) depending on $\gamma$. The Gr\"uneisen parameter, $\gamma$, can be determined as follows \cite{gru}:
\begin{equation}
\gamma=\frac{3(1+\nu)}{2(2-3\nu)}.
\end{equation}
According to Julian \cite{julian}, the coefficient $R(\gamma)$  can be calculated as:
\begin{equation}
R(\gamma)=\frac{5.720\times 10^7\times0.849}{2\times(1-0.514/{\gamma}+0.228/\gamma^2}
\end{equation}

\begin{figure*}
\centering
\includegraphics[width=27.25pc]{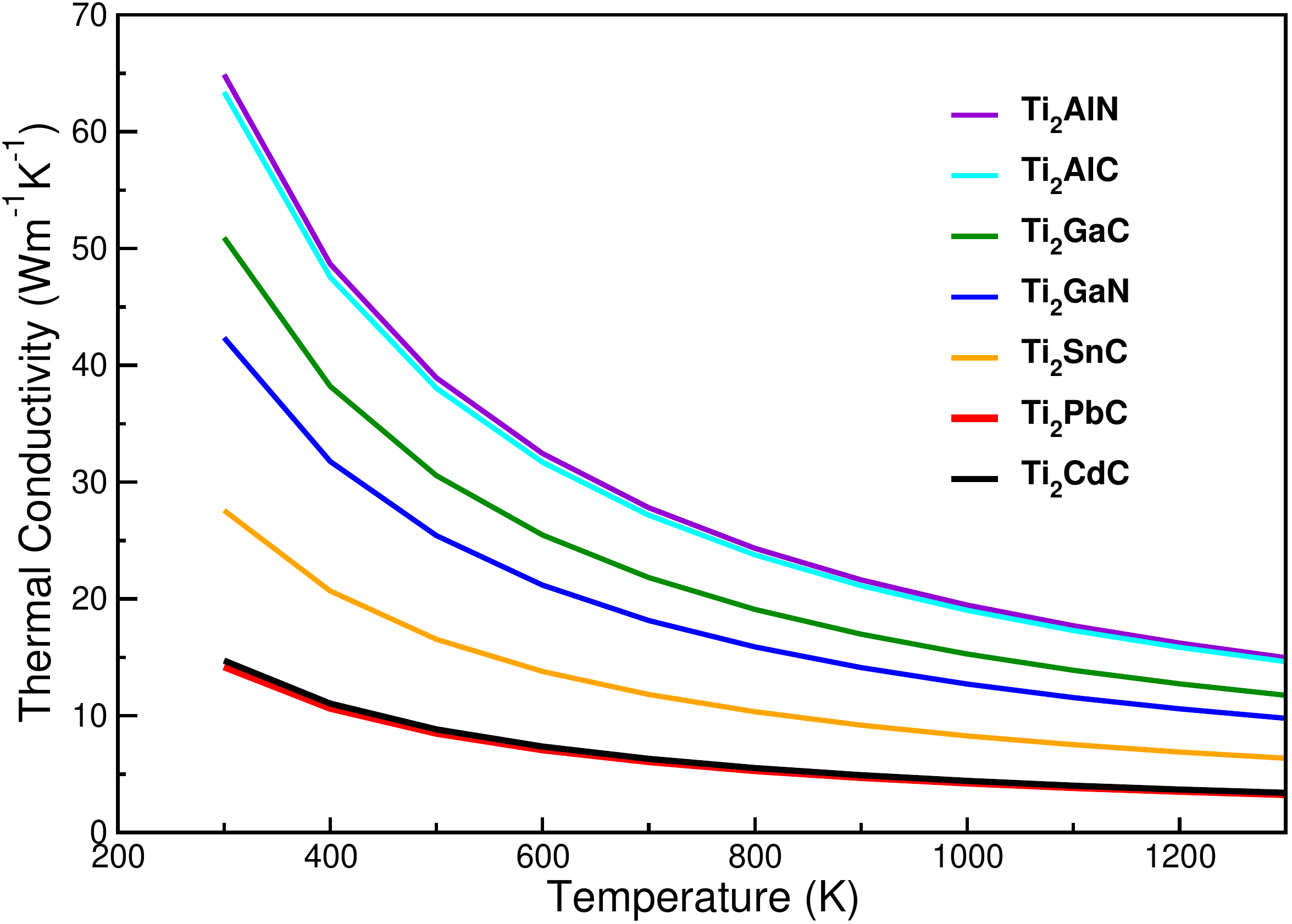}
\caption{\label{Fig:th}Calculated temperature dependent phonon thermal conductivity for the 211 MAX phases calculated using VASP.}
\end{figure*}
\begin{table*}[!h]
\caption{\label{tab:kmin}$\kappa_{min}$ of Ti$_2$AlC, Ti$_2$AlN, Ti$_2$GaC, Ti$_2$GaN, Ti$_2$PbC, Ti$_2$CdC and Ti$_2$SnC MAX phases calculated using VASP compared with the available theoretical data.}
\centering
 \begin{tabular}{c c c c c c c c} 
 \hline
& Ti$_2$AlN& Ti$_2$AlC&Ti$_2$GaC&Ti$_2$GaN& Ti$_2$SnC&Ti$_2$CdC&Ti$_2$PbC\\
\hline
$\kappa_{min}$& 1.41& 1.35&1.24&1.18&0.94&0.84&0.73\\
& -& 1.373 \cite{chandra}&-&-&-&-&-\\
\hline \hline
 \end{tabular}
 \end{table*}
The obtained results for the lattice thermal conductivities of MAX phases are shown in Figure \ref{Fig:th}. From our obtained data, it should be noted that the MAX phases which contains Al-atoms have the highest $\kappa_{ph}$. This is expected since Al-atom is the lightest as compared to  Ga, Cd, Sn and Pb atoms and hence Al scatters less phonons resulting into a higher phonon thermal conductivity. Moreover, the $\kappa_{ph}$ for Ti$_2$AlC of 14.6 W/mK obtained at 1300 K is comparable to the one obtained by Chandra et al. (14.71 W/mK) \cite{chandra} but slightly lower than the experimental value of  16 W/mK \cite{barsoum}. This behaviour is consistent with the results obtained by Chandra et al. \cite{chandra}. It is also evident from Figure \ref{Fig:th} that the estimated $\kappa_{ph}$ of Ti$_2$PbC is the lowest among the studied MAX phases. This is expected since Pb is heavier than Al, Ga, Cd and Sn and thus the $\kappa_{ph}$ of Ti$_2$PbC should be small as compared to the $\kappa_{ph}$ of other considered MAX phases.
\subsection{Phonon properties}
The dynamical property of atoms in the harmonic approximation is obtained by solving eigenvalue problem of dynamical matrix $D(\textbf{q})$ \cite{Ziman},
\begin{equation}
D(\textbf{q})\textbf{e}_{\textbf{q}j}=\omega^2\textbf{e}_{\textbf{q}j},
\end{equation}
where $\textbf{q}$ is the wave vector, and $j$ is the band index, $\omega_{\textbf{q}j}$ and $\textbf{e}_{\textbf{q}j}$ are respectively the phonon frequency and polarization vector of the phonon mode labeled by a set $\{\textbf{q},j\}$.
\par
\begin{figure*}[!h]
\begin{minipage}{9.5pc}
\includegraphics[width=9.5pc]{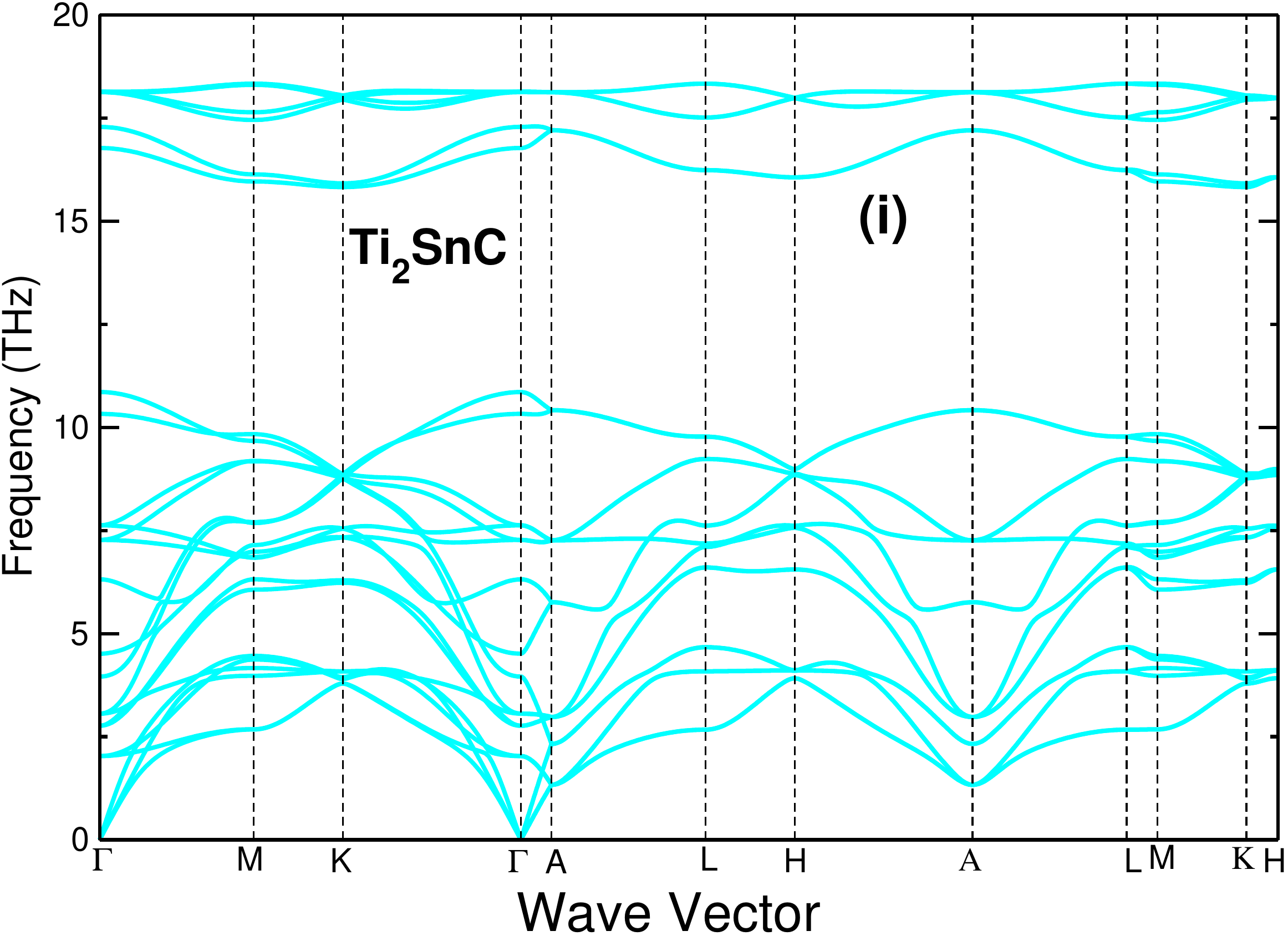}
\end{minipage}\hspace{0.05pc}%
\begin{minipage}{9.15pc}
\includegraphics[width=9.15pc]{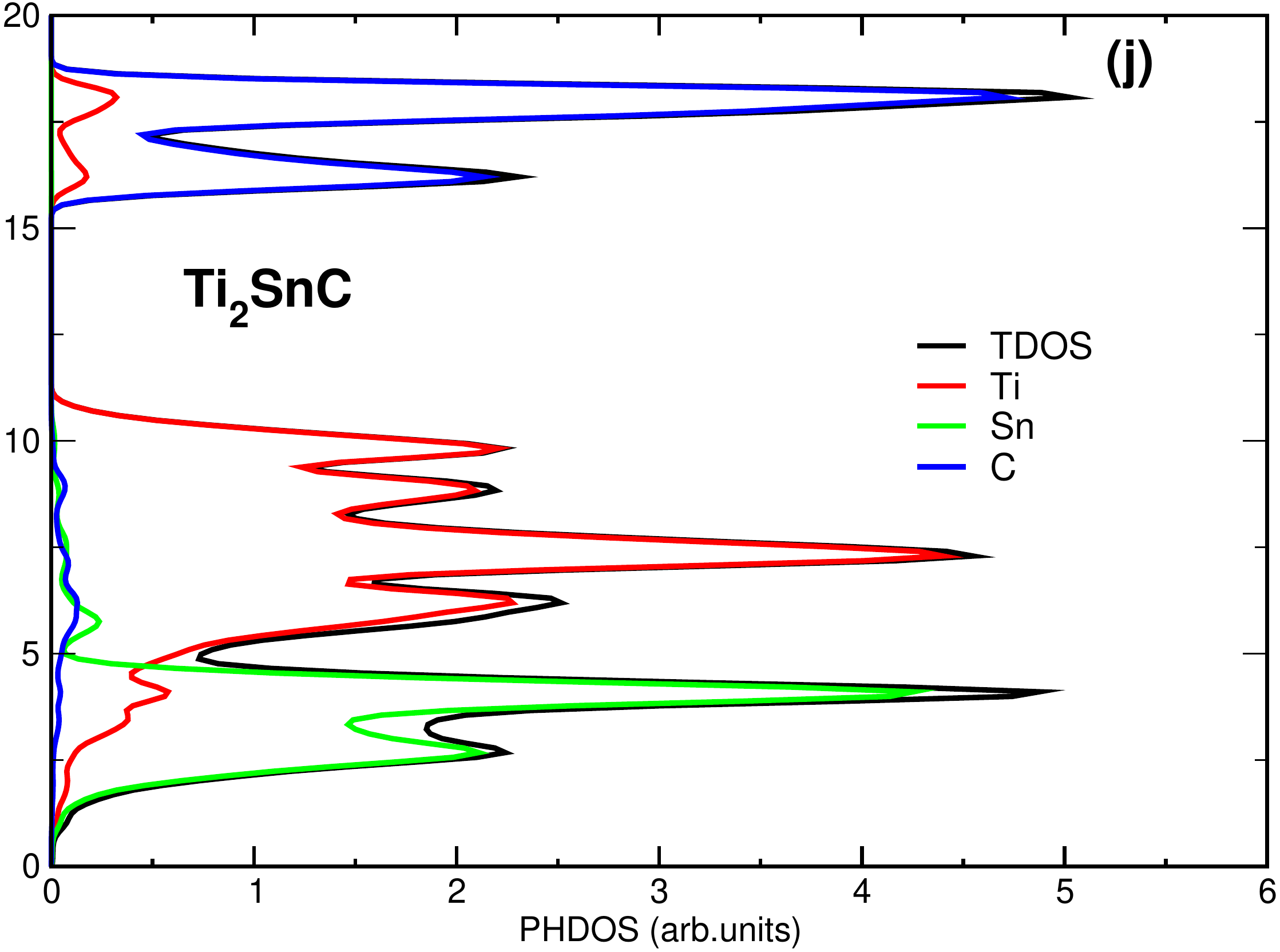}
\end{minipage}\hspace{0.3pc}
 \begin{minipage}{9.5pc}
\includegraphics[width=9.65pc]{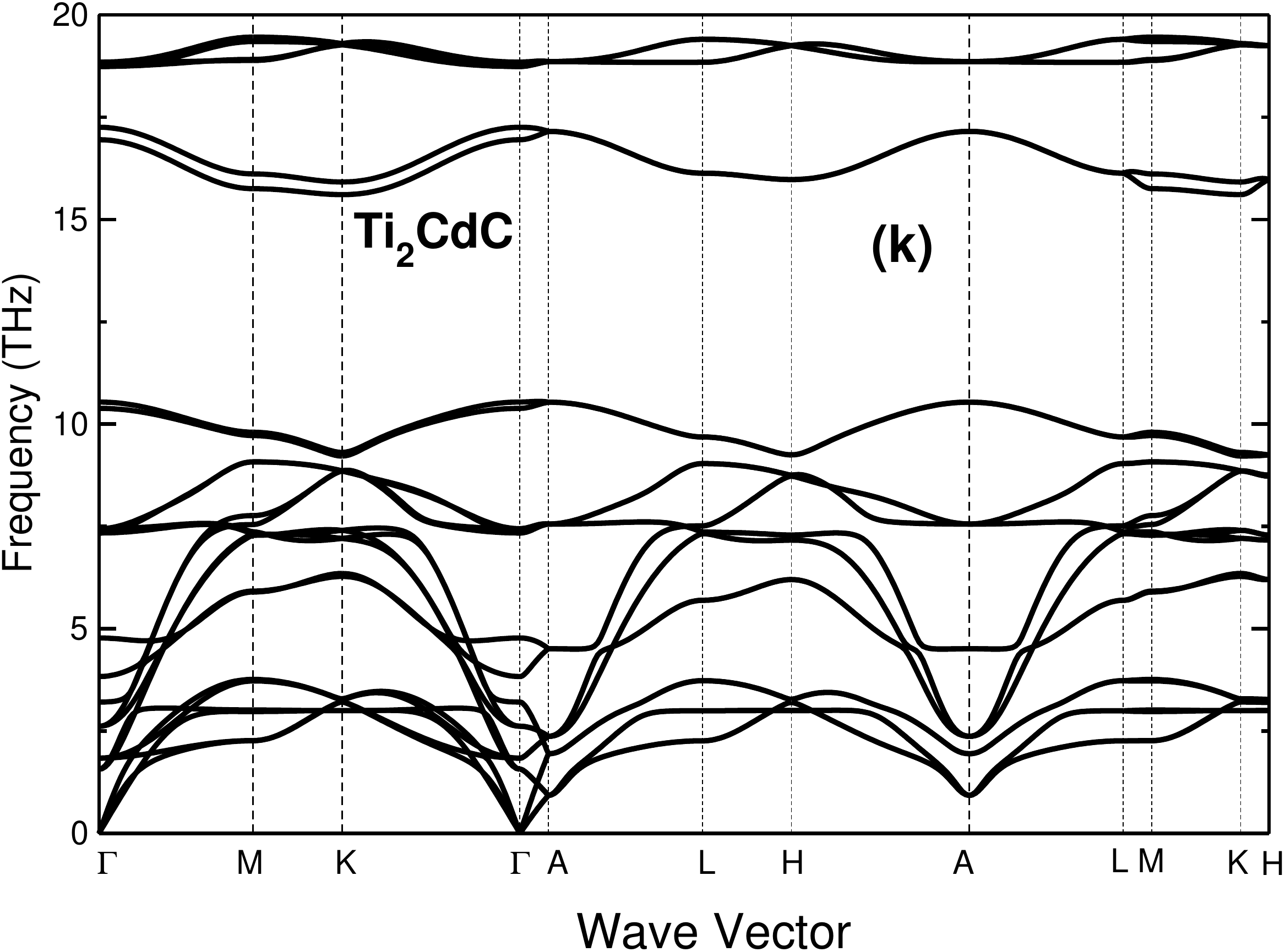}
\end{minipage}\hspace{0.15pc}%
\begin{minipage}{9.5pc}
\includegraphics[width=9.5pc]{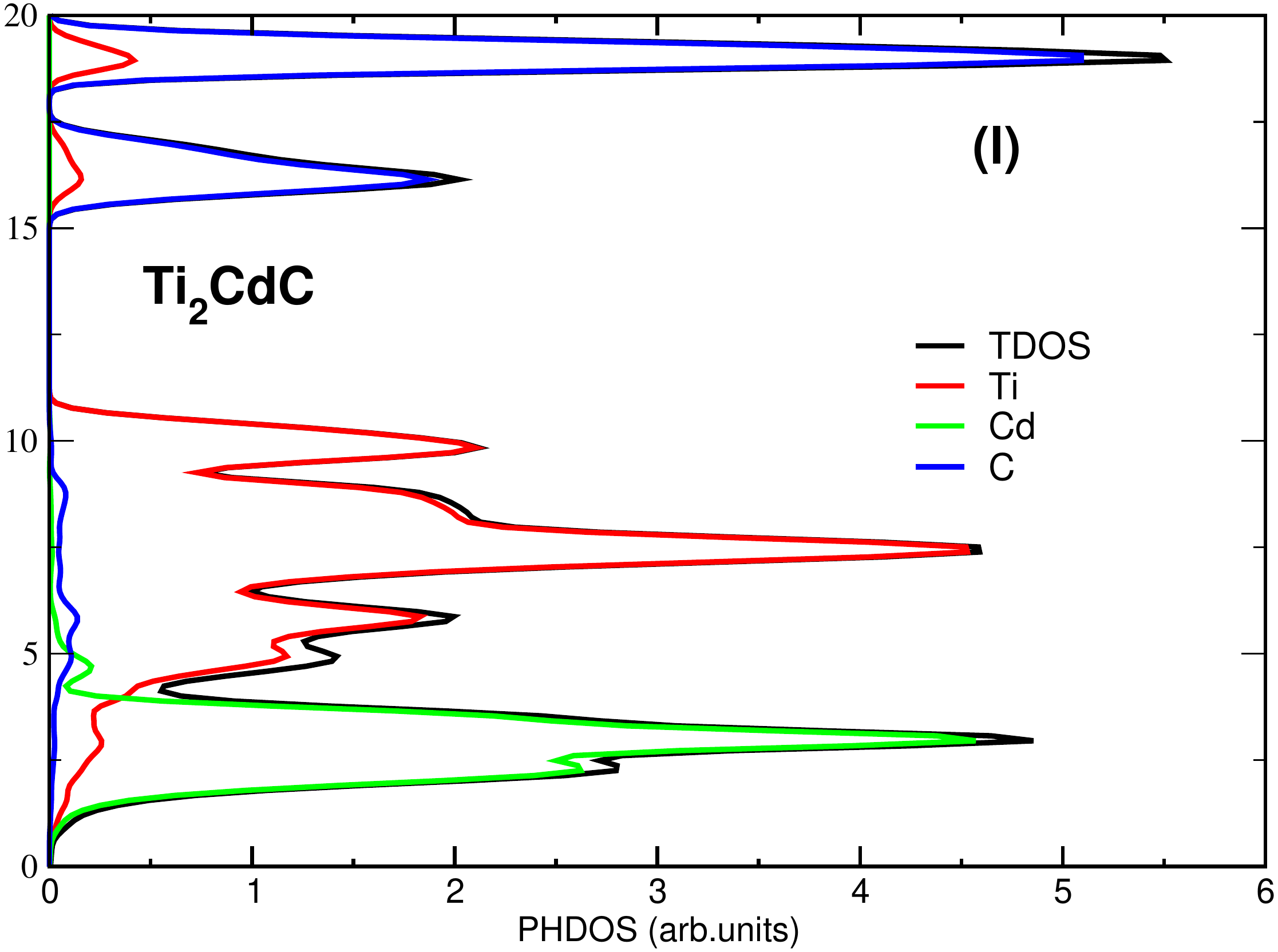}
\end{minipage}\hspace{0.05pc}
 \begin{minipage}{9.5pc}
\includegraphics[width=9.65pc]{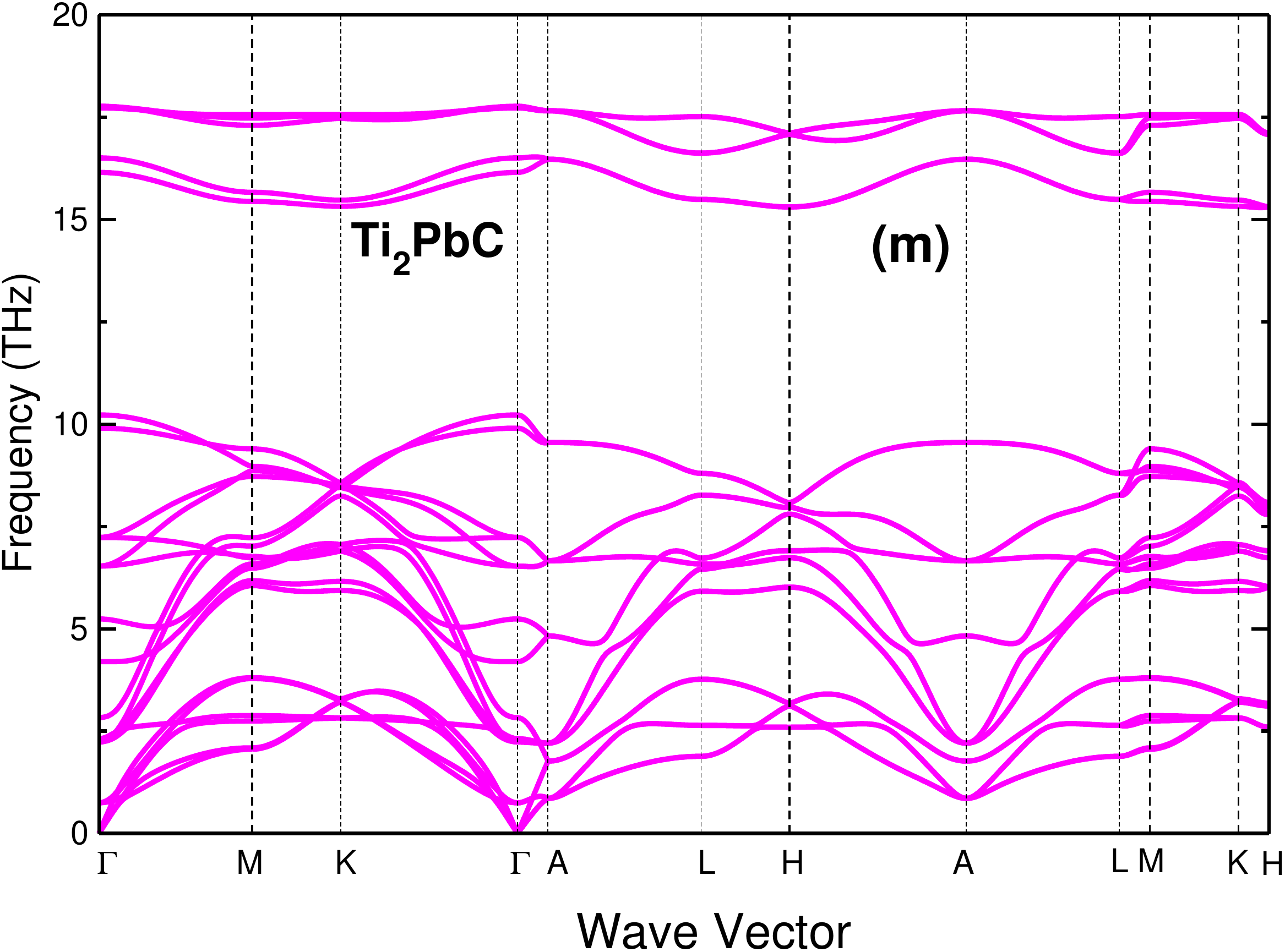}
\end{minipage}\hspace{0.15pc}%
\begin{minipage}{9.5pc}
\includegraphics[width=9.5pc]{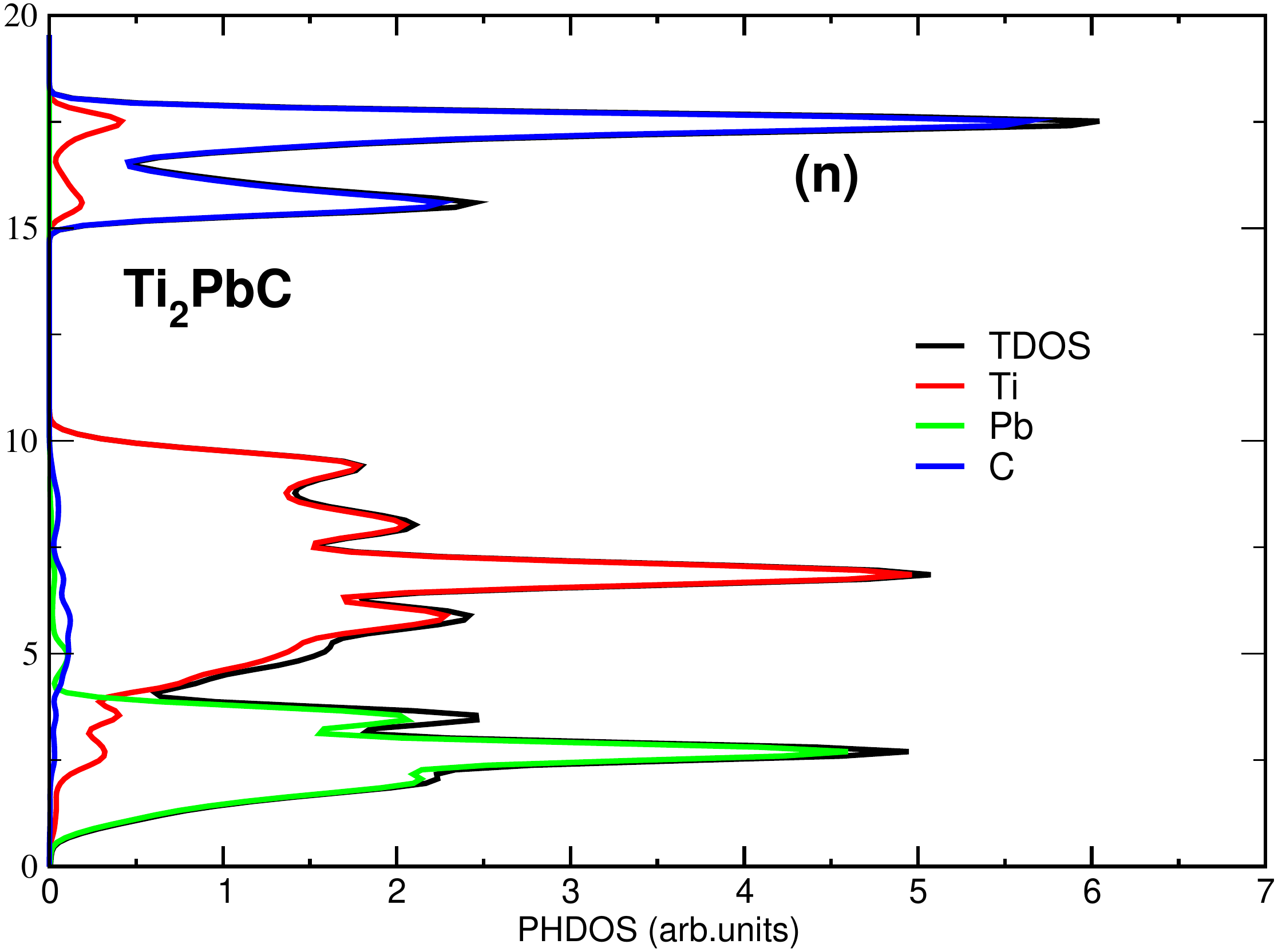}
\end{minipage}\hspace{0.05pc}
\caption{\label{Fig:asph}Phonon dispersion curves and phonon density of
states (PHDOS) for Ti$_2$AlC (a, b), Ti$_2$AlN (c, d), Ti$_2$GaN (e, f), Ti$_2$GaC (g, h), Ti$_2$SnC (i, j), Ti$_2$CdC (k, l) and Ti$_2$PbC (m, n) respectively, calculated using VASP.}
\end{figure*}

 The calculated phonon dispersion curves and the phonon density of states (PHDOS) curves along several high symmetry lines in Brillouin zone have been calculated and are shown in Figures \ref{Fig:asph} (a-n). From Figures \ref{Fig:asph} (a-n) it is clear that all the phonon frequencies are positive in the Brillouin zone, indicating that the structures of materials under study are dynamically stable. As expected the phonon dispersion curves of the three structures contain 24 phonon modes, including three acoustic and 21 optical branches. The phonon spectra of all the studied materials contain three regions separated by two energy gaps as depicted in Figure \ref{Fig:asph} (a-n). From the PHDOS [see Figures \ref{Fig:asph} (a-n)], the acoustic modes of MAX phases are mainly from the vibrations of Al, Ga, Sn, Cd and Pb, the transition metal, Ti, contribute dominantly at low-frequency optical modes while the contribution to the high-frequency optical modes are mainly from the lightest atom i.e N or C-atoms. 
\begin{figure*}[!h]
\begin{minipage}{9.5pc}
\includegraphics[width=9.5pc]{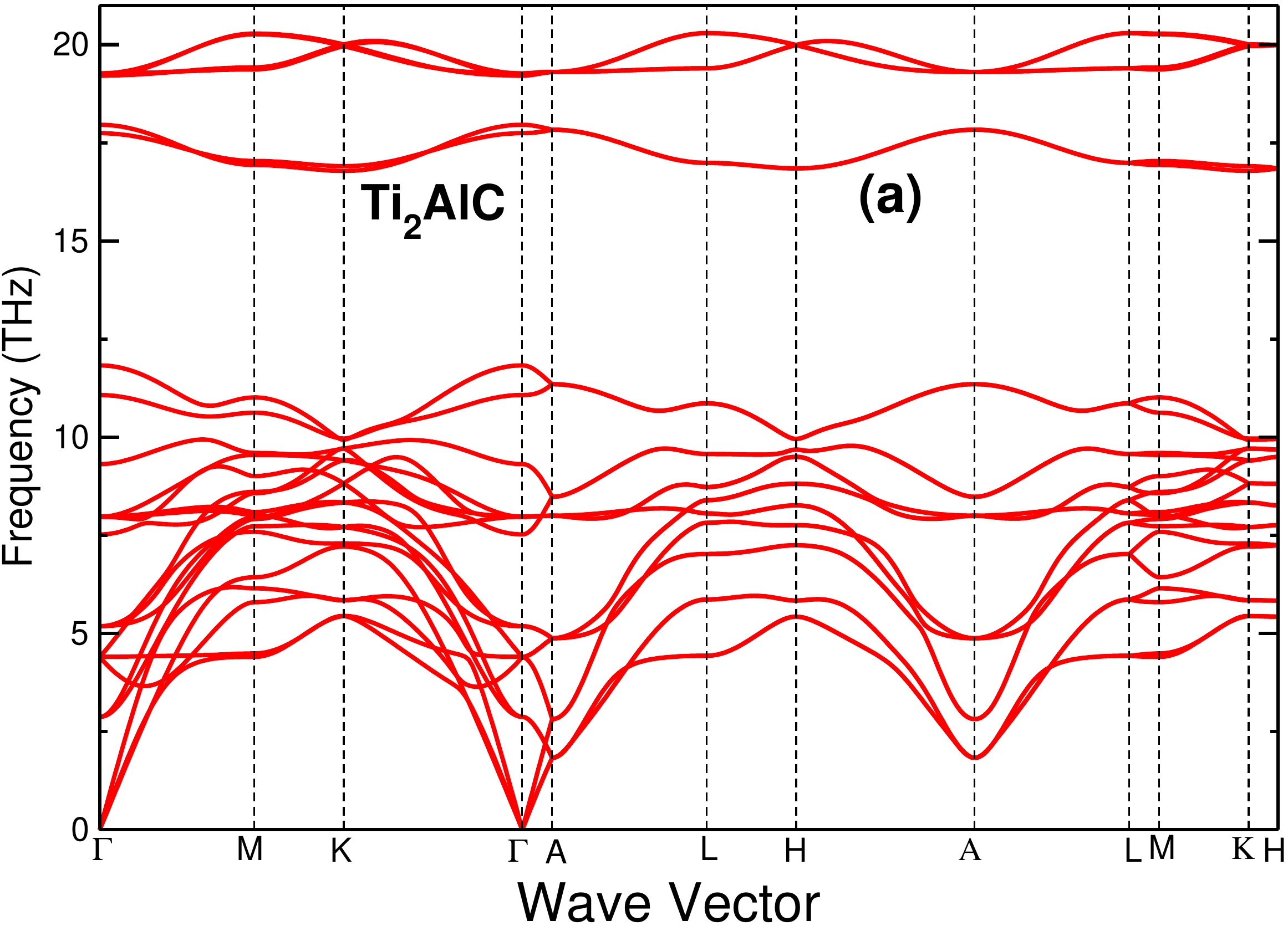}
\end{minipage}\hspace{0.05pc}%
\begin{minipage}{9.15pc}
\includegraphics[width=9.15pc]{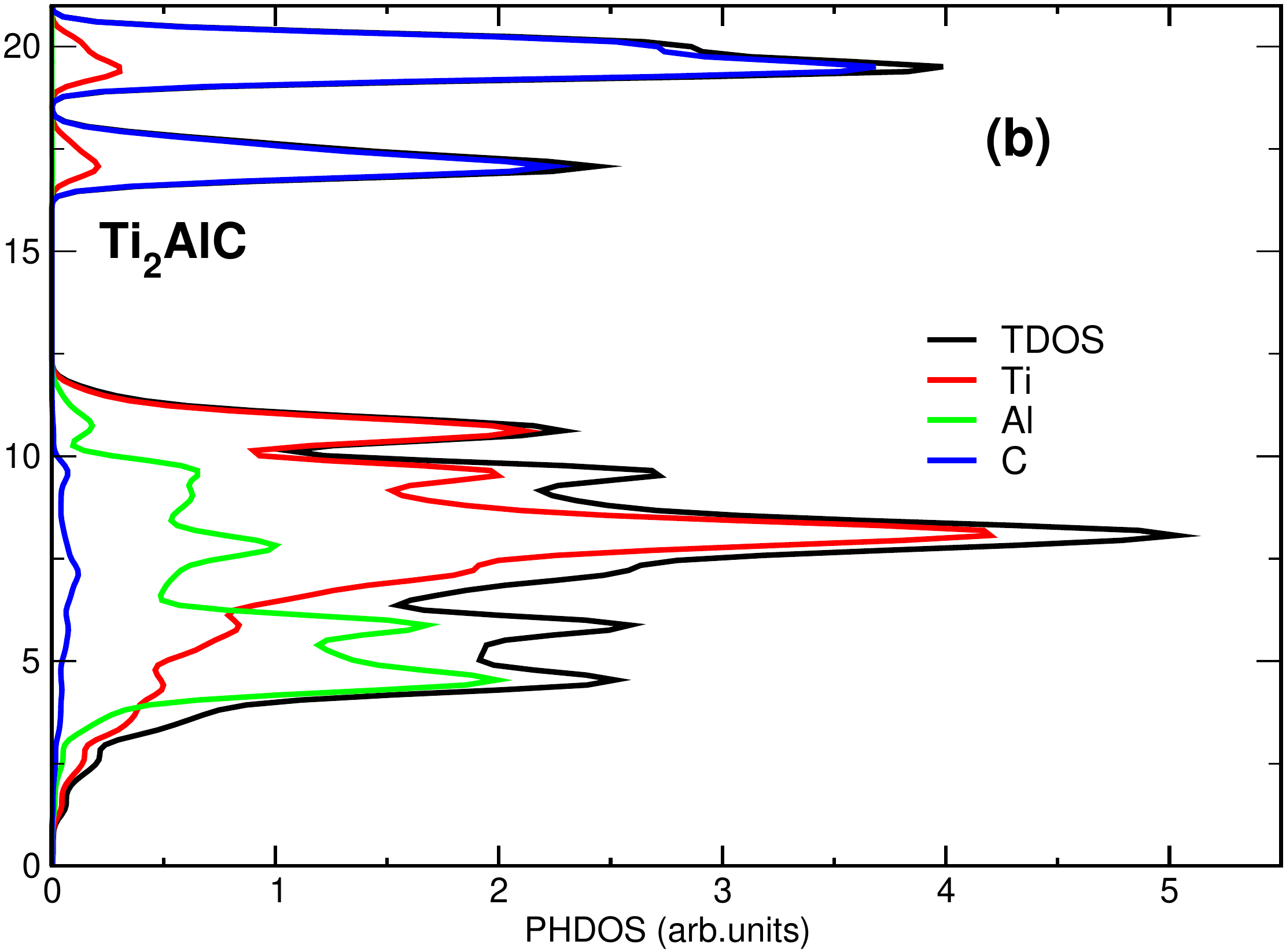}
\end{minipage}\hspace{-0.2pc}
 \begin{minipage}{9.5pc}
\includegraphics[width=9.65pc]{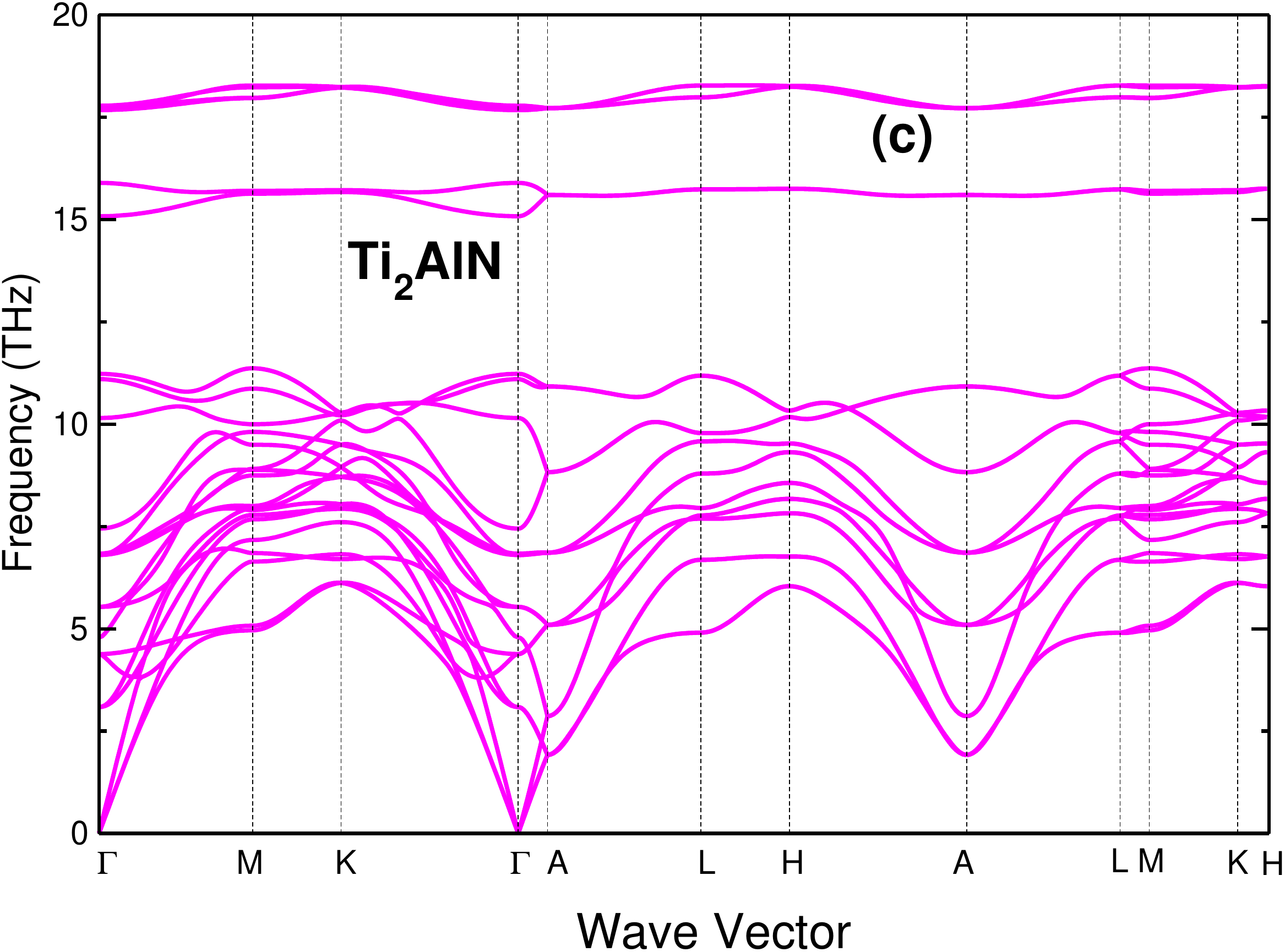}
\end{minipage}\hspace{0.15pc}%
\begin{minipage}{9.5pc}
\includegraphics[width=9.5pc]{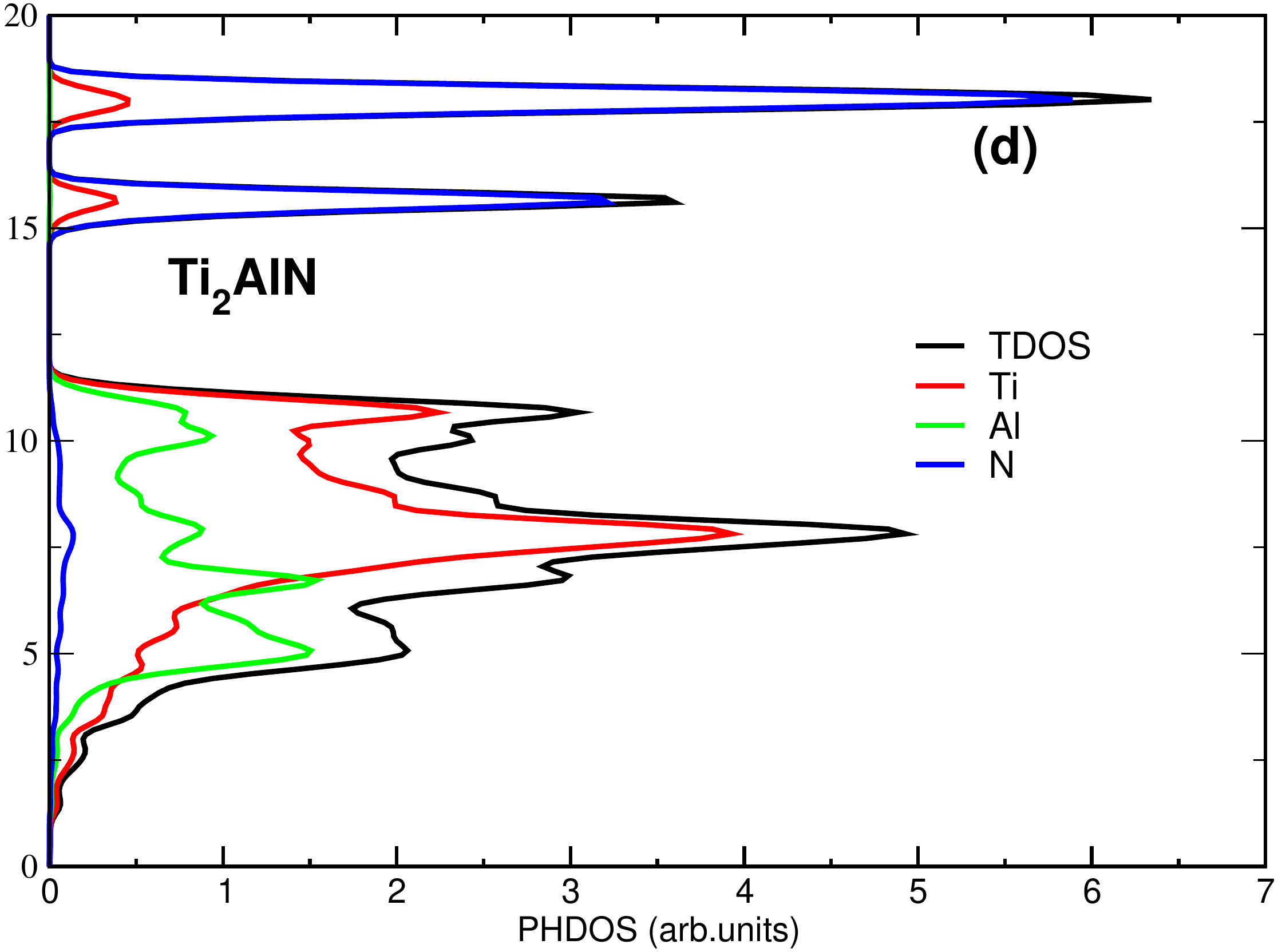}
\end{minipage}\hspace{0.05pc}
\begin{minipage}{9.5pc}
\includegraphics[width=9.65pc]{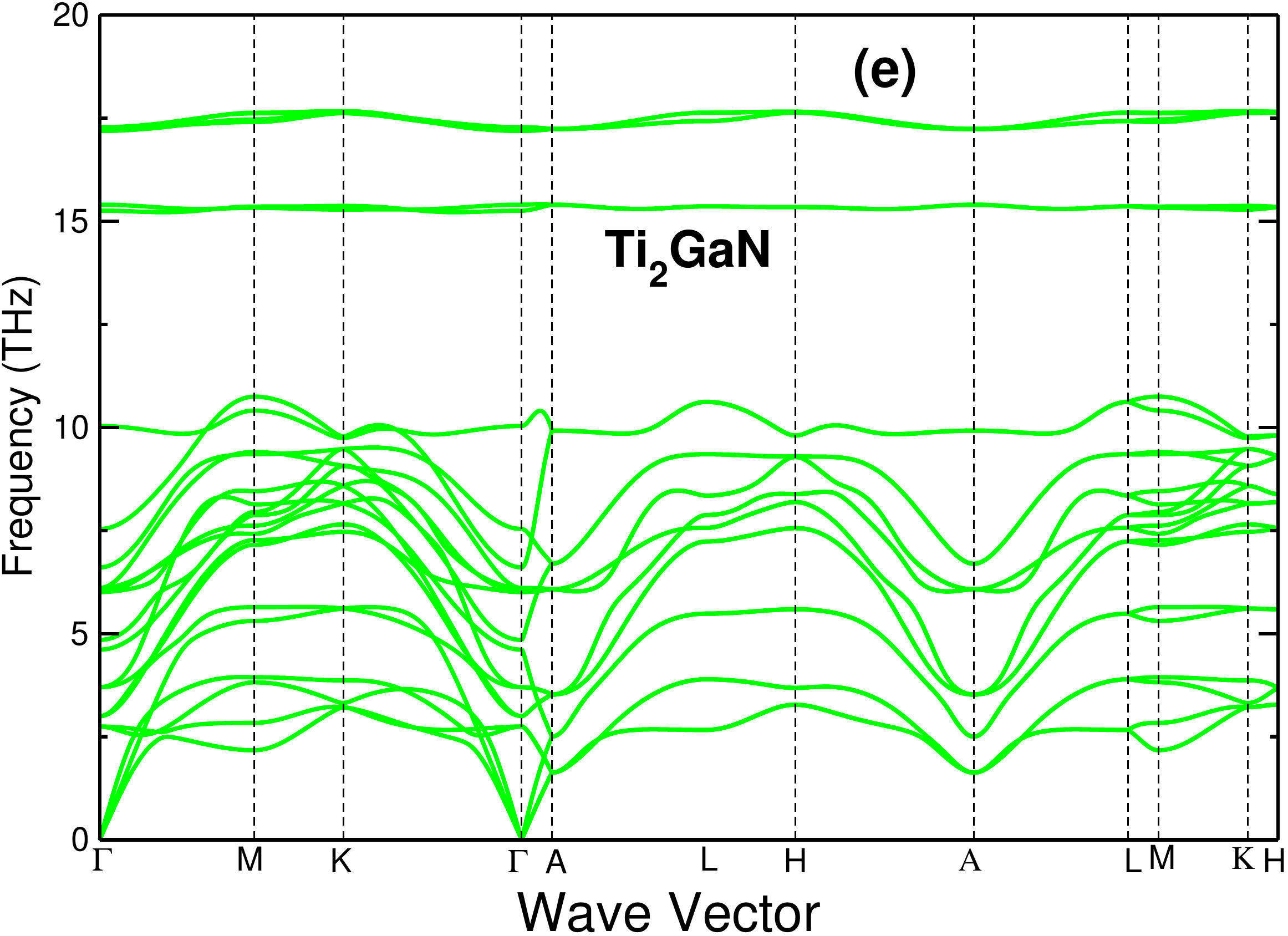}
\end{minipage}\hspace{0.15pc}%
\begin{minipage}{9.5pc}
\includegraphics[width=9.5pc]{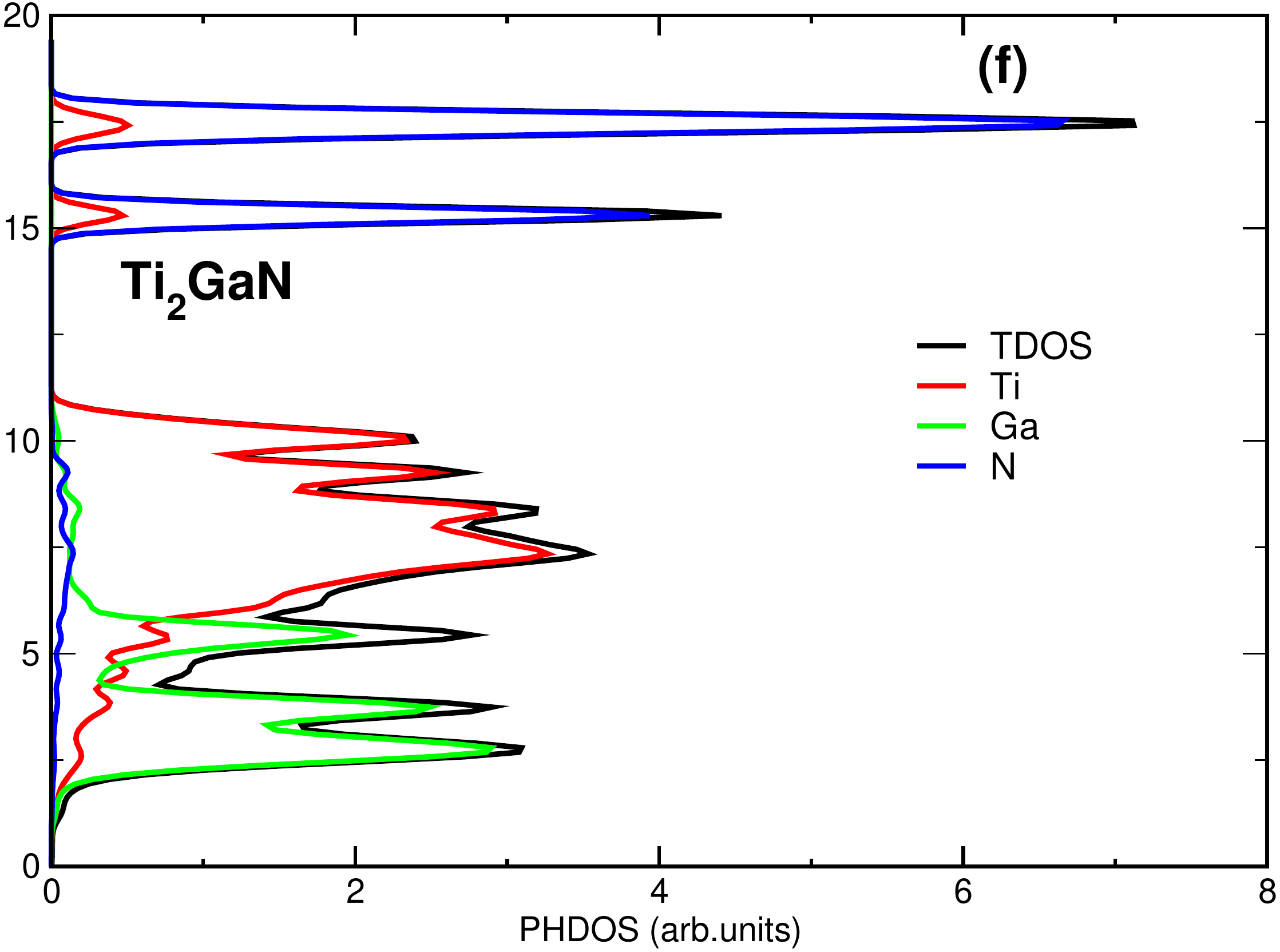}
\end{minipage}\hspace{0.05pc}
\begin{minipage}{9.5pc}
\includegraphics[width=9.5pc]{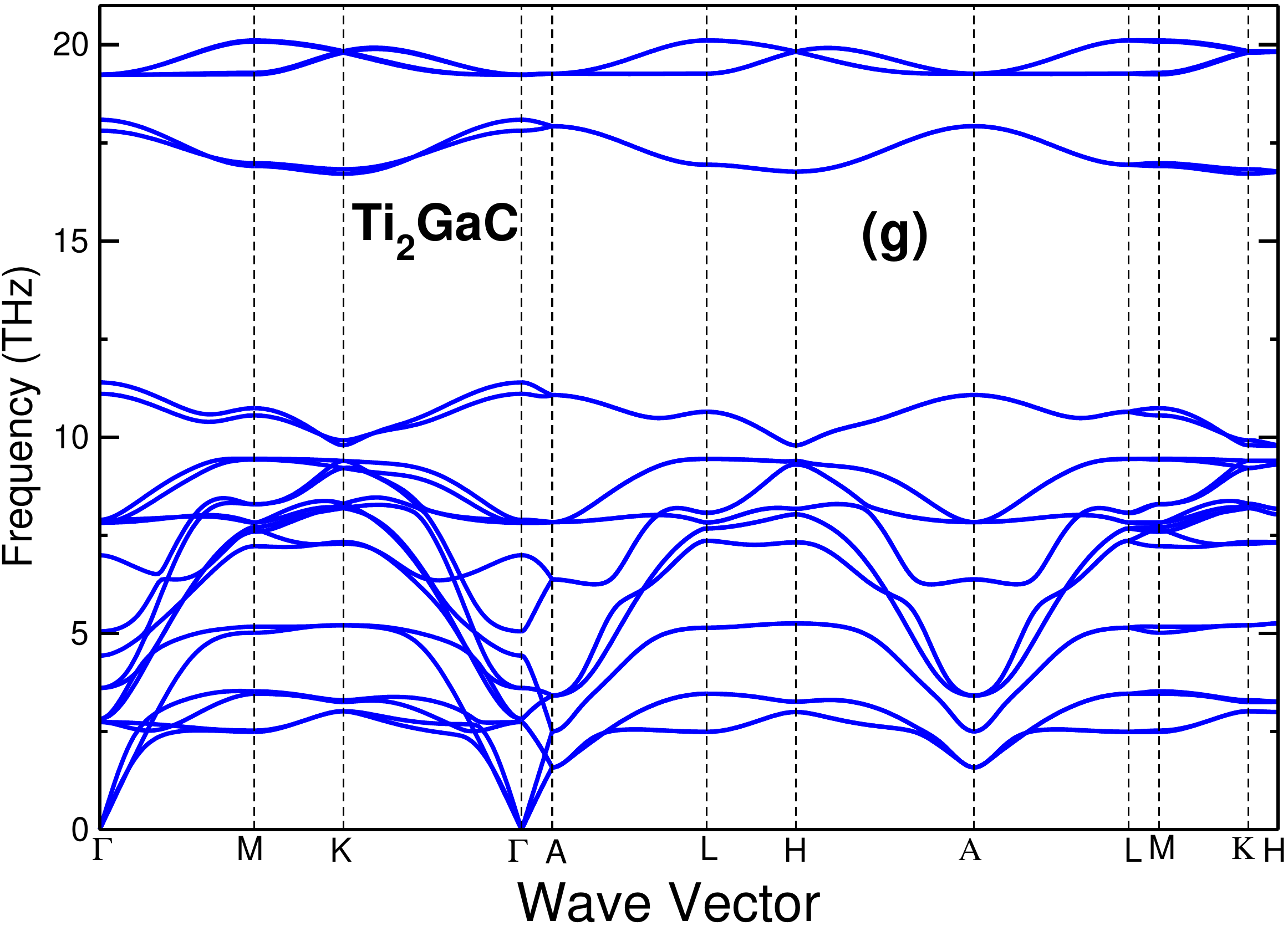}
\end{minipage}\hspace{0.05pc}%
\begin{minipage}{9.15pc}
\includegraphics[width=9.15pc]{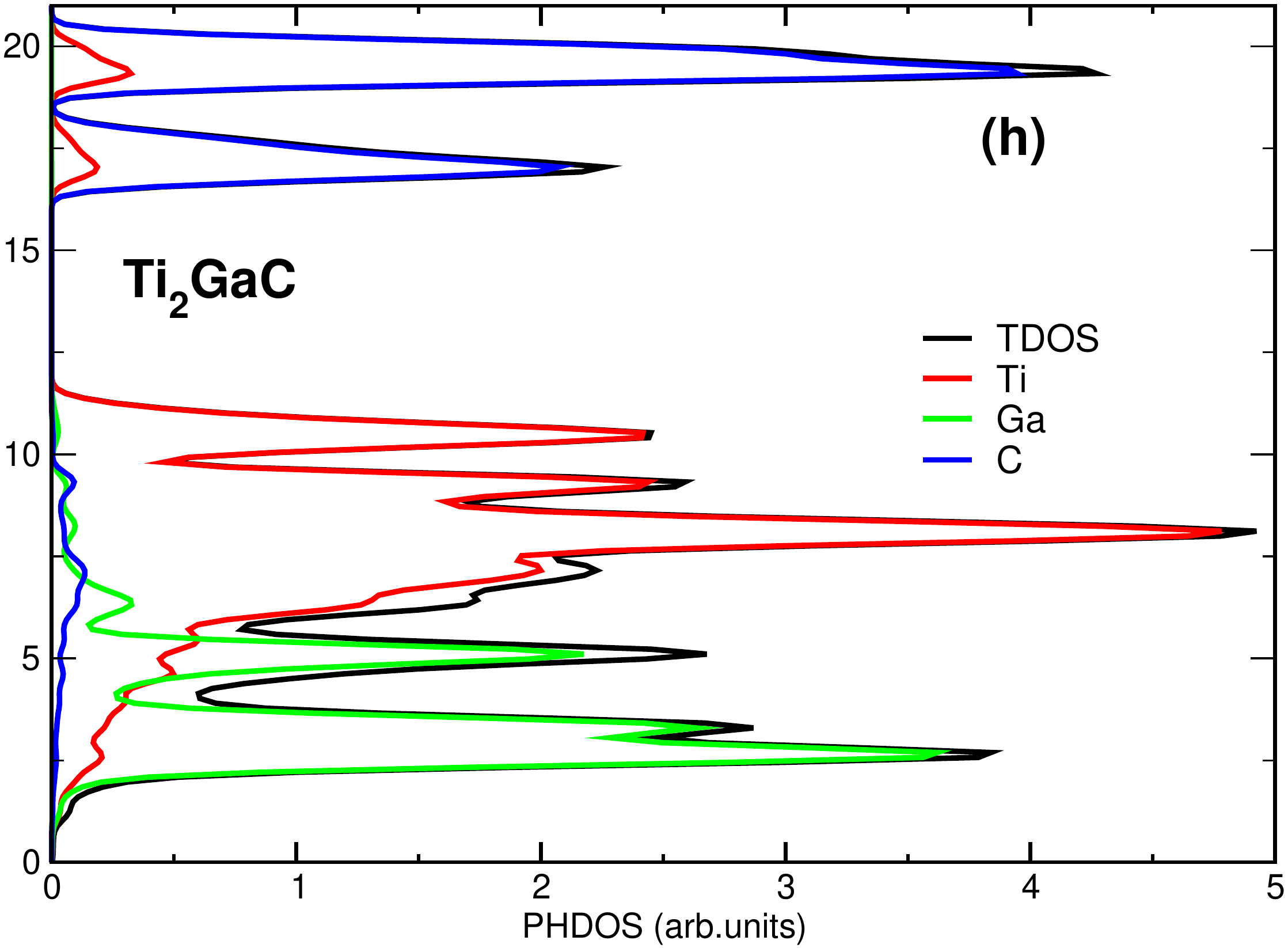}
\end{minipage}\hspace{0.3pc}
\end{figure*} 
\subsection{Thermodynamic properties}
\begin{figure}[!h]
\centering
\includegraphics[width=16.25pc]{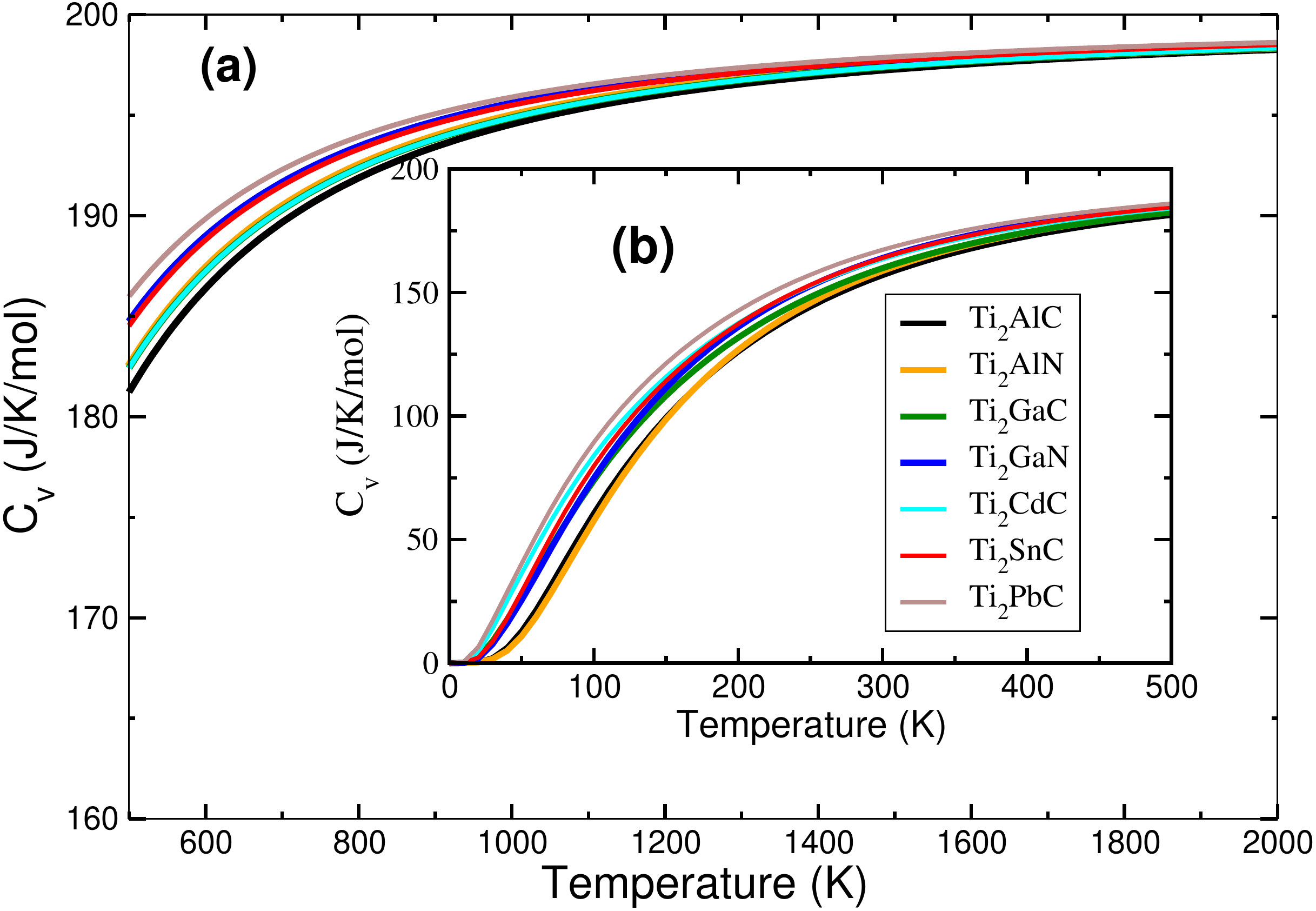}
\caption{\label{Fig:t}Heat capacities of the MAX phases at temperature range (a) 500 K - 2000 K and (b) 0 - 500 K calculated using VASP.}
\end{figure}

The results of the phonon density of states can be used to calculate the lattice heat capacity ($C_v$) as function of temperature. The estimated heat capacities at constant volume as functions of temperature are shown in Figures \ref{Fig:t} in the temperature range from 0 K to 2000 K. Figure \ref{Fig:t} shows that the specific heat, $C_v$, of all the studied materials follows the Debye model which is proportional to $T^3$, as expected \cite{debb}.
\par It is clear that at temperature range between 0 and 300 K, heat capacity increases steadily as the temperature increases. It is also found that the Dulong-Petit law is recovered at high temperatures. The difference in $C_v$ is most prominent at lower temperatures as seen in inset Figure \ref{Fig:t} (b).  From the inset Figure \ref{Fig:t} (b), we noted that the trend of the specific heat capacity at constant volume at T $<$ 250 K is: Ti$_2$AlN $<$ Ti$_2$AlC $<$ Ti$_2$GaC $<$ Ti$_2$GaN $<$ Ti$_2$SnC $<$ Ti$_2$CdC $<$ Ti$_2$PbC.

\section{Conclusion}
We have investigated the structural, mechanical, lattice dynamics properties as well as the approximate thermal conductivity values of the MAX phases by employing the first-principles DFT calculations using QE and VASP codes. We have calculated the independent elastic constants, bulk modulus, shear modulus, Young's modulus, and elastic anisotropy factor. The compounds are shown to be mechanically stable, elastically anisotropic, and ductile in nature. The calculated elastic constants are found to obey the mechanical stability conditions. The dynamical stability of the MAX compounds are confirmed using phonon dispersion curves. The thermodynamic properties such as specific heat capacity is evaluated using the phonon density of states. We have also calculated the lattice thermal conductivity of studied MAX phases using the Slack's model and found that Ti$_2$AlN possesses the highest value while Ti$_2$PbC has the smallest. The minimum thermal conductivity of these MAX phases have also been estimated using Clarke's formula and show that Ti$_2$AlN has the highest value
of $\kappa_{min}$ due to its high value of average sound velocity. 

\section*{Acknowledgments}
We wish to acknowledge support from the Centre for High Performing Computing, South Africa.
\section*{Author contribution statement}

\end{document}